\documentclass[10pt,longnamesfirst,preprint2]{aastex}
\usepackage{amsmath}
\received{2006 December~12}
\revised{2007 April~27}
\accepted{2007 April~30}
\paperid{71086}
\shorttitle{Radio Observations of $\tau$~Boo}
\shortauthors{Lazio \& Farrell}

\newcommand{\tauBoo}{\objectname[]{$\tau$~Boo}}
\newcommand{\mjybm}{\mbox{mJy~beam${}^{-1}$}}

\begin{document}
\title{Magnetospheric Emissions from the Planet Orbiting \tauBoo: A
Multi-Epoch Search}

\author{T.~Joseph~W.~Lazio}
\affil{Naval Research Laboratory, 4555 Overlook Ave.~\hbox{SW}, Washington, DC
	20375-5351}
\email{Joseph.Lazio@nrl.navy.mil}

\and 

\author{W.~M.~Farrell}
\affil{NASA Goddard Space Flight Center, Code~695, Greenbelt, MD 20771}
\email{william.m.farrell@gsfc.nasa.gov}

\begin{abstract}
All of the solar system gas giants produce electron cyclotron masers,
driven by the solar wind impinging on their magnetospheres.
Extrapolating to the planet orbiting \tauBoo, various authors have
predicted that it may be within the detection limits of the 4-meter
wavelength (74~MHz) system on the Very Large Array.  This paper
reports three epochs of observations of \tauBoo.  In no epoch do we
detect the planet; various means of determining the upper limit to the
emission yield single-epoch limits ranging from~135 to~300~mJy.  We
develop a likelihood method for multi-epoch observations and use it to
constrain various radiation properties of the planet.  Assuming that
the planet does radiate at our observation wavelength, its typical
luminosity must be less than about~$10^{16}$~\hbox{W}, unless its
radiation is highly beamed into a solid angle $\Omega \ll 1$~sr.
While within the range of luminosities predicted by various authors
for this planet, this value is lower than recent estimates which
attempt to take into account the stellar wind of \tauBoo\ using the
known properties of the star itself.  Electron cyclotron maser
emission from solar systems planets is beamed, but with characteristic
solid angles of approximately 1~sr illuminated.  Future
long-wavelength instruments (e.g., the Long Wavelength Array and the
Low Frequency Array) must be able to make typical flux density
measurements on short time scales ($\approx 15$~min.)  of
approximately 25~mJy in order to improve these constraints
significantly.
\end{abstract}


\section{Introduction}\label{sec:intro}

The star \tauBoo\ is an F6IV star located 15.6~pc away \citep{plk+97}
that is orbited by a planet with a minimum mass of~4.14~M${}_J$
(Jovian masses) orbiting in a 3.3~day period \citep[semi-major axis
of~0.047~\hbox{AU},][]{bmwhs97}.

By analogy to the ``magnetic planets'' in the solar system
(\objectname[]{Earth}, \objectname[]{Jupiter}, \objectname[]{Saturn},
\objectname[]{Uranus}, and \objectname[]{Neptune}), there have been
various predictions that Jovian-mass extrasolar planets should also
emit intense cyclotron maser emission at radio wavelengths
\citep{zqr+97,fdz99,ztrr01,lfdghjh04,s04,gmmr05,z06,z07}.  Extrapolating empirical relationships based on
the solar system planets, it is possible to make quantitative
predictions for both the characteristic emission wavelength and radio
luminosity for an extrasolar planet
\citep{fdz99,lfdghjh04,s04,gmmr05}.

For the planet orbiting \tauBoo, these predictions are that its
characteristic emission wavelength should be between about~5 and~7
meters (45 and~60~MHz) and that its radio luminosity would result in a
flux density at the \objectname[]{Earth} between roughly 1 and~250~mJy
\citep{lfdghjh04,s04,gmmr05}.  The lower flux density estimates result
from treating \tauBoo\ as effectively a solar twin, while the higher
estimates take into account its (higher) level of stellar activity
resulting from it being younger than the Sun \citep{s04,gmmr05}.  In
addition, for the solar system planets, variations within the level of
solar activity can amplify the cyclotron maser emission process,
producing radio luminosities (and therefore flux densities) 1--2
orders of magnitude above the nominal level.

Indirect evidence for extrasolar planetary magnetic fields comes in
the form of modulations in the \ion{Ca}{2} H and~K lines of the stars
\objectname[HD]{HD~179949} and \objectname[]{$\upsilon$~And},
modulations that are in phase with the orbital periods of their
planets with the smallest semi-major axes \citep{swbgk05}.  Though
they monitored \tauBoo, no similar modulations were seen.  
\cite{swbgk05} suggest that the \ion{Ca}{2} line modulations result from
energy transport related to the relative velocity between planet and
the stellar magnetosphere.  Any such \ion{Ca}{2} line modulations
for \tauBoo\ would then be suppressed because the star's rotation
period is comparable to the planet's orbital period \citep[$\approx
3$~d,][]{cdsba07}.  Suggestively, though, the polarization
observations of \cite{cdsba07} do suggest a complex surface magnetic
field topology for \tauBoo, consistent with a possible interaction
with the planet's magnetic field.

For the solar system planets, the cyclotron maser emission is fairly
wideband with $\lambda/\Delta\lambda \sim 2$ ($\Delta\nu/\nu \sim
1/2$).  The Very Large Array (VLA) is equipped with a 4-meter
wavelength (74~MHz) receiving system.  Images can be made with the VLA
in its more extended configurations (A and~B) with rms noise levels of
approximately 100~\mjybm.  Thus, the radio emission from the planet
may extend to wavelengths of~4~meters, and the planet may be
detectable with current instrumentation.

This paper reports three epochs of~74~MHz observations of \tauBoo\
with the \hbox{VLA}.  In \S\ref{sec:observe} we describe the
observations; in \S\ref{sec:analyze} we analyze our observations both
from the standpoint of the individual epochs as well as taken
collectively and we make suggestions about the methodology for
observations with future radio telescopes; and in \S\ref{sec:conclude}
we present our conclusions.

\section{Observations}\label{sec:observe}

Our observations were conducted on three epochs with the VLA in its
more extended configurations.  Table~\ref{tab:log} summarizes various
observational details.  The use of the extended VLA configurations
provided angular resolutions comparable to or better than 1\arcmin;
such resolution is required to reduce the impact of source confusion.
Dual circular polarization was recorded at all epochs, and the total
bandwidth was 1.56~MHz, centered at~73.8~MHz.

\begin{deluxetable}{lcccc}
\tablecaption{74~MHz Observational Log\label{tab:log}}
\tablewidth{0pc}
\tablehead{
 \colhead{Epoch} & \colhead{Configuration}
	& \colhead{Duration} & \colhead{Beam} & \colhead{Image Noise Level} \\
	         &
	& \colhead{(min.)}   & \colhead{(\arcsec)} & \colhead{(\mjybm)}}
\startdata
1999 June~8       & D$\to$A & 281           & $30 \times 29$ & 120 \\
2001 January~19   & B       & \phantom{1}75 & $80 \times 80$ & 110 \\
2003 September~12 & BnA     & \phantom{1}66 & $32 \times 21$ & 100 \\
\enddata
\end{deluxetable}

Post-processing of~4-m wavelength VLA data uses procedures similar to
those at shorter wavelengths, although certain details differ.  The
source \objectname[]{Cygnus~A} served as both the bandpass, flux
density, and visibility phase calibrator.  For two of the epochs (1999
June~8 and 2003 September~12), visibility phase calibration was
performed in much the same manner as at shorter wavelengths, with the
phases determined from \objectname[]{Cygnus~A} transferred to the
\tauBoo\ data.  Several iterations of hybrid mapping (imaging and
self-calibration) then ensued.  For the middle epoch, the data were
acquired as part of the VLA Low-frequency Sky
Survey \citep[VLSS,][]{clk+07}.  For these data, small ``postage stamp'' images of
bright sources from the NRAO VLA Sky Survey \citep{ccgyptb98} within
the field of view were made every 1~min.  The positions of the NVSS
sources are determined at~20~cm, a wavelength at which the ionosphere
should present only minor perturbations.  The apparent positions
at~4-meter wavelength were compared with the known positions
from~20~cm.  Low-order Zernicke polynomials were used to model the set
of position offsets and thereby infer the phase corrections to be
applied to the data.  After applying these phase corrections, the full
data set was imaged.

Two significant differences for the post-processing were the impact of
radio frequency interference (RFI) and the large fields of view. In
order to combat RFI, the data were acquired with a much higher
spectral resolution than used for imaging. Excision of potential RFI
is performed on a per-baseline basis for each visibility spectrum.
For two of the epochs (1999 June~8 and 2003 September~12), RFI was
identified and excised manually; for the second epoch (2001
January~19), RFI was identified and excised on an automated basis by
fitting a linear function to each 10-s visibility spectrum and
removing outlier channels.  The large field of view (11\arcdeg\
at~74~MHz) mean that the sky cannot be approximated as flat. In order
to approach the thermal noise limit, we used a polyhedron algorithm
\citep{cp92} in which the sky is approximated by many two-dimensional
``facets.''

\begin{figure}[tb]
\epsscale{0.8}
%\rotatebox{-90}{\plotone{1999.ps}}
\rotatebox{-90}{\plotone{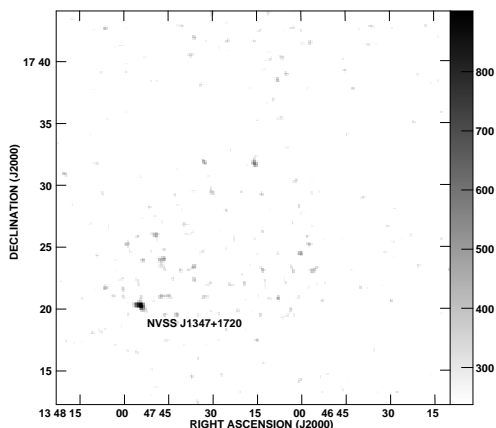}}
\vspace*{-3ex}
\caption[]{The field around \tauBoo\ at~74~MHz for the epoch 1999
June~8.  The grey scale flux range is linear between~240 ($2\sigma$)
and~900~\mjybm.  The source in the lower left is
\objectname[NVSS]{NVSS~J1347$+$1720} and is used to assess the
accuracy of our astrometry.}
\label{fig:epoch1}
\end{figure}

\begin{figure}[tb]
%\rotatebox{-90}{\plotone{2001.ps}}
\rotatebox{-90}{\plotone{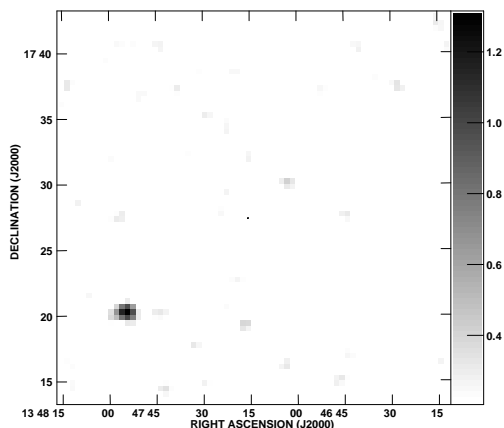}}
\vspace*{-3ex}
\caption[]{As for Figure~\ref{fig:epoch1} for the epoch 2001
January~19.  The grey scale flux range is linear between~220
($2\sigma$) and~1300~\mjybm.}
\label{fig:epoch2}
\end{figure}

\begin{figure}[tb]
%\rotatebox{-90}{\plotone{2003.ps}}
\rotatebox{-90}{\plotone{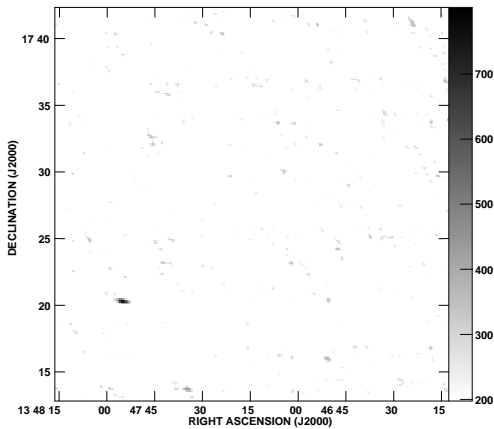}}
\vspace*{-3ex}
\caption[]{As for Figure~\ref{fig:epoch1} for the epoch 2003 September~12.  The grey scale flux range is linear between~200 ($2\sigma$) and~800~\mjybm.}
\label{fig:epoch3}
\end{figure}

Figures~\ref{fig:epoch1}--\ref{fig:epoch3} present 74~MHz images of the field
around \tauBoo\ at the three epochs.

\section{Discussion and Analysis}\label{sec:analyze}

We have used a number of different methods in order to assess the
limits on the presence of any emission at the location of \tauBoo.  We
begin with an assessment of the astrometric accuracy of the images.

\subsection{Astrometric Accuracy}\label{sec:astrometry}

Ionospheric phase fluctuations result in refractive shifts in the
apparent positions of sources within the images.  In all of the images
(Figures~\ref{fig:epoch1}--\ref{fig:epoch3}) is the source
\objectname[NVSS]{NVSS~J1347$+$1720}.  The position of this source is
determined at a shorter wavelength (20~cm or~1400~MHz) at which the
ionospheric-induced position shifts are unimportant.  For each epoch,
we have fit an elliptical gaussian to
\objectname[NVSS]{NVSS~J1347$+$1720} and determined the position from
that fit.  Uncertainties, in either right ascension or declination,
are between~0\farcs3 and~6\arcsec.

In addition, \tauBoo\
has a measured proper motion of $\mu_\alpha = -480.34 \pm
0.66$~mas~yr${}^{-1}$ and $\mu_\delta = 54.18 \pm
0.40$~mas~yr${}^{-1}$, measured in the epoch 1991.25 \citep{plk+97}.
Over the decade between the epoch of the proper motion determination
and the measurements reported here, we expect \tauBoo\ to have moved
by no more than about~5\arcsec.  Adding these uncertainties in
quadrature, we expect that the position of \tauBoo\ in these images
should be uncertain by no more than about~8\arcsec.  The size of the
beam (point spread function) ranges from about~25\arcsec\ (epochs~1
and~3) to~80\arcsec\ (epoch~2).  We conclude that our astrometry is
accurate to a fraction of a beam width.

\subsection{Single-Epoch Limits on the Planetary Radio Emission}\label{sec:limits}

In no epoch do we detect statistically significant emission at the
location of \tauBoo.  Our first estimate for the upper limit to any
emission from the planet orbiting \tauBoo\ is obtained from the rms
noise level in each image (Table~\ref{tab:log}).  Because our
astrometry is accurate to better than a beam width, we set an upper
limit of~$2.5\sigma$ at each epoch.  The resulting upper limits are
250--300~mJy.

As a second means for setting an upper limit, we use the brightest
pixel within a beam centered on the position of \tauBoo\ to estimate
the flux density of any possible radio emission from its planet.  This
estimate takes into account the background level determined by the
mean brightness in a region surrounding the central beam.  Upper
limits determined in this manner range from~135 to~270~mJy.

As a final means for setting an upper limit, we have co-added
(``stacked'' or superposed epoch analysis) the three images, a
technique has been used with great success to find weak sources in
diverse data sets (e.g., sources contributing to the hard X-ray
background, \citealt*{wfb+05}; intergalactic stars in galaxy clusters,
\citealt*{zwsb05}).  Experience with 74~MHz images shows that the rms
noise level in an image produced from the sum of~$N$ images is
$\sqrt{N}$ lower, as expected if the noise in the images is gaussian
random noise.  Obviously, by co-adding images, we will be less
sensitive to a rare, large enhancement in the planetary radio emission
caused by a temporary increase in the stellar wind flux, but we will
be more sensitive to the nominal flux density, particularly for the
higher estimates for the planet's flux density \citep[e.g.,][]{s04}.
The ($1\sigma$) noise level in the co-added image is 65~\mjybm, a
value consistent with that expected if the noise in all of the images
is gaussian.  We do not detect any source at the position of \tauBoo\
at the level of~165~mJy ($2.5\sigma$).

Table~\ref{tab:limits} summarizes these limits.  At any given epoch,
the upper limits on radio emission at~74~MHz from the planet orbiting
\tauBoo\ range from an optimistic 135~mJy to, more conservatively,
300~mJy.  On average, the planet's flux density is not larger than
165~mJy.  We convert these flux densities to luminosities assuming
that the emission is broadband ($37\,\mathrm{MHz} = 1/2$ of the
observing frequency) and that the planet radiates into a solid angle
of~1~sr (see below).

\begin{deluxetable}{lccc}
\tablecaption{74~MHz \protect\tauBoo\ Observational Limits\label{tab:limits}}
\tablewidth{0pc}
\tablehead{
 \colhead{Method} & \colhead{Epoch} & \multicolumn{2}{c}{Limit} \\
                  &                 & \colhead{(mJy)} & \colhead{($\times 10^{16}$~W)}}
\startdata
rms noise level & 1999 June~8       & 300 & 2.6 \\
                & 2001 January~19   & 275 & 2.4 \\
                & 2003 September~12 & 250 & 2.1 \\
\\					       
					       
brightest pixel & 1999 June~8       & 270 & 2.3 \\
                & 2001 January~19   & 134 & 1.1 \\
                & 2003 September~12 & 189 & 1.6 \\
\\					       
					       
stacked image   & \ldots            & 165 & 1.4 \\

\enddata
\tablecomments{Luminosity limits adopt a distance of~15.6~pc and
assume an emission bandwidth of~37~MHz ($= 74\,\mathrm{MHz}/2$).}
\end{deluxetable}

This planet does not transit its host star.  While the
planetary magnetosphere might be large, the typical emission altitude
for an electron cyclotron maser for solar system planets is
approximately 1--3 planetary radii.  Thus, these limits should apply
regardless of orbital phase.

\subsection{Likelihood Estimates for Multi-Epoch Planetary Radio
 	Observations}\label{sec:likeli}

Earlier work by \cite{bdl00}, \cite{fdlbz03}, \cite{rzr04}, and
\cite{fldbz04} (and
\S\ref{sec:limits}) reported non-detections from a single observation
of a planet.  (In addition, \citealt*{yse77} and \citealt*{wdb86}
conducted blind searches for extrasolar planetary radio emission, but
it is not clear that they observed any star now known to be orbited by
a planet, and we believe that they observed each star for only a
single epoch.)  In addition, \tauBoo\ has been observed at~150~MHz
with the Giant Metrewave Radio Telescope \citep[GMRT,][]{mwcklnz06},
the results from which will be reported elsewhere.  Here we have
reported non-detections from multiple observations of \tauBoo.  Both
in the context of these observations as well as from the standpoint of
future observations, to what extent can multiple non-detections of a
planet be used to place constraints on its radio emission?

The expected flux density for the radio emission from an extrasolar
planet is \citep{fdz99,lfdghjh04}
\begin{equation}
S = \frac{P_{\mathrm{rad}}}{\Delta\nu\Omega D^2},
\label{eqn:fluxdensity0}
\end{equation}
where $P_{\mathrm{rad}}$ is the luminosity or radiated power from the
planet, $\Delta\nu$ is the emission bandwidth of the radiation,
$\Omega$ is the solid angle into which the radiation is beamed, and
$D$ is the distance of the planet (or its host star) from the Sun.  In
turn, the luminosity or radiated power and the emission bandwidth can
be related to various planetary properties (e.g., mass and rotation
rate of the planet).  The standard practice has been to use empirical
laws from the solar system to make predictions for $P_{\mathrm{rad}}$
and~$\Delta\nu$.

Planetary radio emission has a characteristic wavelength~$\lambda_c$
or frequency~$\nu_c$, which is related to the cyclotron frequency at
the lowest altitude at which emission is able to escape
\citep{fdz99}.  In turn, this characteristic wavelength is 
related to the magnetic moment or the magnetic field strength at the
planet's surface.  The emission is typically broadband, with
$\lambda_c/\Delta\lambda =
\nu_c/\Delta\nu \sim 2$.  In making predictions, it has been assumed
that extrasolar planetary radio emission is comparably broadband, and,
more importantly, that any observational searches would be carried out
at a wavelength near $\lambda_c$ (frequency near~$\nu_c$).  From an
observational standpoint, we shall incorporate a factor to take into
account the possibility that a search may not have been conducted at
an optimum wavelength.  Guided by the spectrum of
\objectname[]{Jupiter's} emission, we use
\begin{equation}
f_\nu(\nu, \nu_c) =
 \begin{cases}
 {1}, & \nu < 2\nu_c; \\
 {0}, & \nu > 2\nu_c.
 \end{cases}
\label{eqn:frequency}
\end{equation}
Although the step function nature of $f_\nu(\nu, \nu_c)$ may seem
artificial, \objectname[]{Jupiter's} spectrum does cutoff sharply
around~40~MHz, which is approximately $2\nu_c$.

Combining equations~(\ref{eqn:fluxdensity0})
and~(\ref{eqn:frequency}), and making the assumption that $\Delta\nu
\sim \nu_c/2$, the expected flux density of an extrasolar planet when
observed at a frequency~$\nu$ is 
\begin{equation}
S(P_{\mathrm{rad}}, \nu_c, \Omega; D, \nu)
 = \frac{2}{\nu_c}\frac{P_{\mathrm{rad}}}{\Omega D^2}f_\nu(\nu, \nu_c).
\label{eqn:fluxdensity}
\end{equation}
We take $P_{\mathrm{rad}}$, $\nu_c$ ($\lambda_c$), and~$\Omega$ to be
planetary parameters with the objective of constraining them from
observation.  The distance~$D$ to the planet (or its host star) is
typically known from other means.

Our methodology for searching for radio emission from an extrasolar
planet has been to utilize a radio interferometer to make images of
the field surrounding the planet.  In the absence of a source, the
pixels in a thermal-noise limited image from a radio interferometer
have a zero-mean normal distribution\footnote{
The assumption of a zero-mean distribution depends upon the image
having been made without a so-called ``zero-spacing'' flux density,
that is, without a measurement of the visibility function at a 
spatial frequency of zero wavelengths.  This is the case for the
images we analyze here.}
with a variance of~$\sigma^2$, so that the probability density
function (pdf) of obtaining a pixel with a noise intensity between~$N$
and~$N+dN$ is
\begin{equation}
f_N(N) = \frac{1}{\sigma\sqrt{2\pi}}e^{-N^2/2\sigma^2}.
\label{eqn:noise}
\end{equation}

We adopt a signal model of $I = S + N$ for the intensity at the
location of the planet, where~$N$ is the (thermal) noise in the image
and~$S$ is the flux density contributed by the planet.  We assume that
$S$ is constant over the duration of the observation.  If this is not
the case, then the pdf of~$S$ would have to be incorporated into this
analysis.  In practice, perhaps the best that one could do would be to
use what is known about the radio emission of the giant planets in the
solar system to develop an appropriate pdf.  For the purposes of this
analysis, we shall treat $S$ as a constant, which has the effect that
we will be placing constraints on the mean level of radio emission
from a planet.

A detection occurs if the pixel intensity exceeds some threshold,
$I_t$.  Thus,
\begin{equation}
p_S(I>I_t|S)
 = \int^{\infty}_{I_t} \frac{1}{\sigma\sqrt{2\pi}}\,e^{-(I-S)^2/2\sigma^2}\,dI.
\label{eqn:p}
\end{equation}
For simplicity, we define $x = I/\sigma$ and $s = S/\sigma$.

Equation~(\ref{eqn:fluxdensity}) suggests that small values
of~$\Omega$ would produce large flux densities.  However, small values
of~$\Omega$ also imply that the radiation beam is unlikely to
intersect our line of sight.  We characterize the probability of
intercept as
\begin{equation}
p_\Omega(\Omega) = \frac{\Omega}{2\pi}.
\label{eqn:lpi}
\end{equation}
For the solar system planets, the emission is beamed from both the
northern and southern auroral regions.  We presume that the line of
sight to a planet cannot intercept both regions simultaneously, so
that the the appropriate denominator for this probability is a solid
angle of~$2\pi$~sr.  Clearly more complicated functions are possible,
but we do not believe them warranted at this time.  The full
probability of detection is then
\begin{equation}
p(P_{\mathrm{rad}}, \Omega, \nu_c) = p_Sp_\Omega,
\label{eqn:pfull}
\end{equation}
where we have made explicit the planetary parameters that we seek to determine.

Suppose we have $n$ observations of a planet with $m$ detections, with
$m = 0$ describing the current observational state.  Can we place any
constraints on the factors in equation~(\ref{eqn:fluxdensity})?

Consider first a series of trials in which the probability of
detecting the planet in any single trial~$p$ is the same.  This case
corresponds to one in which the observations are essentially
identical.  Then the probability of detection in the several trials is
given by the binomial probability
\begin{equation}
P(p; m, n) = {n \choose m}p^m(1 - p)^{n-m}.
\label{eqn:binomial}
\end{equation} 
The case of current interest would be that for which $m = 0$,
\begin{equation}
P(p; 0, n) = (1 - p)^n.
\label{eqn:binomial0}
\end{equation}

In any actual observational case, $p$ will likely vary from epoch to
epoch, most likely because the images will have different noise
levels~$\sigma$.  This is certainly the case if the images are
obtained from different instruments (e.g., the VLA vs.\ the Giant
Metrewave Radio Telescope).  Even images from the same instrument can
have different noise levels, though, depending upon the prevalence of
RFI during the observations, the number of antennas used, the duration
of the observation, different elevations at which the source is
observed, etc. (Table~\ref{tab:log}).

We assume that the observations are independent, that is that the
probability of detecting the planet in any given observation is
independent of the other observations.  This assumption is certainly
warranted from the observational standpoint that the noise
level~$\sigma$ is independent from observational epoch to epoch.  Then
the \emph{joint} probability of detection is
\begin{equation}
{\cal P}
 = \prod_{i=1}^N P_i
 = \prod_{i=1}^N {n_i \choose m_i}p_i^{m_i}(1 - p_i)^{n_i-m_i},
\label{eqn:ptotal}
\end{equation}
for which the total number of trials~$n_i$ and number of
detections~$m_i$ are allowed to vary from one set of trials to
another.  We give the full expression here, anticipating that there
may be future observations, potentially at a range of wavelengths
(e.g., VLA vs.\ GMRT vs.\ Long Wavelength Array vs.\ Low Frequency
Array, see below).  For the case presented here, there is a single
trial at each epoch with no detection, $n_i = 1$, $m_i = 0$, and $N =
3$, so
\begin{equation}
{\cal P} = \prod_{i=1}^3 (1 - p_i).
\label{eqn:total}
\end{equation}
For a set of observed threshold intensities~$\{x_t\}$, the likelihood
function is given by equation~(\ref{eqn:total})
\begin{equation}
{\cal L}(\{x_t\} | P_{\mathrm{rad}}, \Omega, \nu_c)
 = \prod_{i=1}^3 [1 - p(x>x_{i,t}|s)]
\label{eqn:ltotal}
\end{equation}
where we have made explicit the parameter dependences entering into $S$.

For the observations reported here, because we have observations at
only one wavelength, we assume that the observing frequency is
sufficiently close to $\nu_c$ that $f_\nu(\nu, \nu_c) = 1$.  In this
case, the likelihood function (equation~\ref{eqn:ltotal}) reduces to a
function of only two parameters, $P_{\mathrm{rad}}$ and~$\Omega$.
Also, in practice, we compute the (base~10) logarithm of the
likelihood.

\begin{figure}[tb]
\epsscale{0.67}
%\rotatebox{-90}{\plotone{likelg.ps}}
\rotatebox{-90}{\plotone{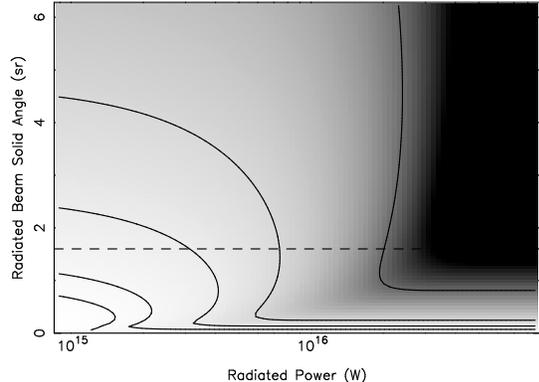}}
\vspace{-2ex}
\caption[]{The (log-)likelihood function, equation~\ref{eqn:ltotal},
for detecting the radio emission from the planet orbiting \tauBoo\ as
a function of its radiated power~$P_{\mathrm{rad}}$ and the solid
angle~$\Omega$ into which the radiation is beamed, assuming that it
does radiate at a wavelength of~4~m.  The gray scale is linear, with
white indicating regions consistent with our non-detections and black
indicating unlikely regions.  The rms noise levels and flux densities
used are listed in Tables~\ref{tab:log} and~\ref{tab:limits}; the
bandwidth of the emission is assumed to be 37~MHz ($=
74\,\mathrm{MHz}/2$).  The maximum occurs in the lower left, near
$P_{\mathrm{rad}} \simeq 10^{15}$~W and $\Omega \simeq 0.2$~sr; the
first contour is at~99\% of the peak, with subsequent contours
at~98\%, 95\%, 90\%, and~67\%.  The horizontal dotted line indicates
the approximate beaming solid angle for \protect\objectname[]{Jupiter}
\citep{zc04}.}
\label{fig:likeli}
\end{figure}

Figure~\ref{fig:likeli} shows the (log-)likelihood function, which we
have computed with the image noise levels listed in
Table~\ref{tab:log} and the brightest pixel limits
(Table~\ref{tab:limits}).  The peak likelihood occurs in the lower 
central left of the plot, near $P_{\mathrm{rad}} \sim 10^{15}$~W and
$\Omega \simeq 0.2$~sr.  The location of this peak reflects our choice
of limits on~$P_{\mathrm{rad}}$, as lower power levels would also
clearly be consistent with our non-detections.  

Two comments on the likelihood function are warranted.  
First, the shape of the allowed region reflects two competing
effects.  From equation~(\ref{eqn:fluxdensity}), the quantities
$P_{\mathrm{rad}}$ and $\Omega$ are degenerate.  The planet could
radiate intensely but be beamed into a narrow solid angle, with a low
probability of detection, or it could have a wide beaming angle but
with only modest luminosity.  The competing effect is that, as the
beaming angle becomes smaller, the probability of it intersecting our
line of sight becomes progressively smaller.

Second, the upper limit on~$P_{\mathrm{rad}}$ in
Figure~\ref{fig:likeli} is motivated by predictions.  A crucial
element of the prediction by \cite{s04} is that the stellar mass loss
rate was predicted on the basis of the star's X-ray flux.  It now
seems that stellar mass loss rates change character at a
characteristic X-ray flux of about $8 \times
10^5$~erg~cm${}^{-1}$~s${}^{-1}$ \citep{wmzlr05}.  Below this value,
stellar mass loss rates are highly correlated with X-ray flux; above
this value, the mass loss rates drop by an order of magnitude and
remain relatively constant with increasing X-ray fluxes.  The X-ray
flux of \tauBoo\ is essentially equal to the value at which stellar
mass loss rates change character.  Thus, its mass loss rate could be
overestimated by as much as an order of magnitude.
\cite{s04} predicts that the planet's cyclotron maser strength should
scale with the mass loss rate as $\dot M^{2/3}$.  Thus, even if it is
the case that the mass loss rate was overestimated by a factor of~10,
the radiated power from the planet could still be a factor of~5 above
that which would be estimated assuming a solar mass loss rate.

For our observations, assuming that the planet does radiate at a
wavelength of~4~m, we conclude that the upper end of the predicted
power levels is allowed only if the beaming angle is $\Omega \ll
1$~sr.  The cyclotron maser emission from the solar system planets is
characterized by solid angles $\Omega \gtrsim 1$~sr, typically emitted
in a hollow cone, with a fairly wide opening angle ($\sim 90\arcdeg$)
and a finite thickness.  Recent analysis of Cassini observations of
\objectname[]{Jupiter} during the spacecraft's cruise phase indicate
that its decameter radiation illuminates a cone of half-width opening
angle of~75\arcdeg\ and a thickness of~15\arcdeg, with an equivalent
solid angle $\Omega \simeq 1.6$~sr \citep{zc04}.  In some cases, for
example the Earth's auroral kilometric radiation, the beaming angle
can be $\Omega \gtrsim \pi$~sr \citep{gg85}.

For beaming angles comparable to those seen in the solar
system ($\Omega \sim 1$~sr), the planet must radiate less than
about~$10^{16}$~W or we would have a reasonable probability of
detecting it.  Luminosities $P_{\mathrm{rad}} < 10^{16}$~W are within
the predicted range \citep{lfdghjh04,s04,gmmr05}, though below the most recent
estimates that attempt to incorporate what is known about the star's
stellar wind strength.  For comparison, \objectname[]{Jupiter's}
nominal radio luminosity is of order $10^{10}$~\hbox{W}.

\cite{bdl00} also subdivided each observation of a single star into
scans as short as 10~s (the shortest nominal integration time provided
by the VLA).  The motivation is that \objectname[]{Jupiter's} emission
is quite variable, so, if the extrasolar planet radio emission is
similar, subdividing the observation offers the possibility of
detecting a ``burst,'' which might be quite short in duration and
``diluted'' if the entire ($\gtrsim 1$~hr) observation is considered.
Though \cite{bdl00} did not observe \tauBoo\ at~4-m wavelength, we
have considered the impact of subdividing the observation.  The
obvious benefit of subdividing is that the effective number of
observations increases; the disadvantage is that the noise level does
as well.

We assume similar observing parameters as obtained in our observing
programs, namely a point-source noise level $\sigma \approx 0.1$~Jy
obtained in~1~hr for observations toward a star like \tauBoo.  We find
that sub-dividing the observation into~10 scans (e.g., each of~5~min.\
duration), which increases the noise level to $\sigma \approx 0.3$~Jy,
leads to essentially no improvement in the constraints that can be
set.

As a final comment on our likelihood method, we anticipate future
observations, either with existing or future instruments (see below)
might be at different wavelengths.  A significant assumption in our
likelihood analysis is that searches are conducted at a common
wavelength and that the planet emits at that wavelength.  In the case
of \objectname[]{Jupiter}, for instance, its electron cyclotron
emission cuts off sharply shortward of approximately 7.5~meters ($\nu
> 40$~MHz).  Presuming that similar processes operate in the
magnetospheres of extrasolar giant planets, observations at different
wavelengths may not be equally constraining.  Similarly, one might
wish to take into account what is known about the beaming angles from
the solar system planets.  It would be relatively straightforward to
incorporate this prior information and extend our likelihood analysis
to a Bayesian formulation.  In that case, appropriate priors would
have to be specified for the emission wavelength, beaming angle, and
radiated power.

\subsection{Future Observations}\label{sec:future}

There are a number of next-generation, long wavelength radio
instruments under development.  Notable among these are the Low
Frequency Array (LOFAR) and the Long Wavelength Array (LWA).  If
deployed as intended, both promise to provide sensitivities at least
an order of magnitude larger than the 4-meter wavelength VLA at
comparable wavelengths ($\lambda > 3$~m or $\nu < 100$~MHz).  Here we
consider whether additional observations with current instrumentation
is preferable to the operation of these future facilities.

We simulated two sets of flux density measurements toward a star like
\tauBoo.  For both measurements the typical flux density measurement
was taken to be a gaussian random variable with a specified mean and
variance and the rms noise level was taken to be a factor of~2.5 times
smaller.  The first set of measurements had noise levels and flux
density limits similar to those measured here, but we considered the
impact of having 15 measurements rather than just 3.  Not detecting
planetary radio emission in such a data set would not improve
significantly on the radiated power constraints (at most by a factor
of a few) but would place increasingly severe constraints on the
beaming angle at large power levels.

\begin{figure}[tb]
\epsscale{0.67}
%\rotatebox{-90}{\plotone{future_fake.ps}}
\rotatebox{-90}{\plotone{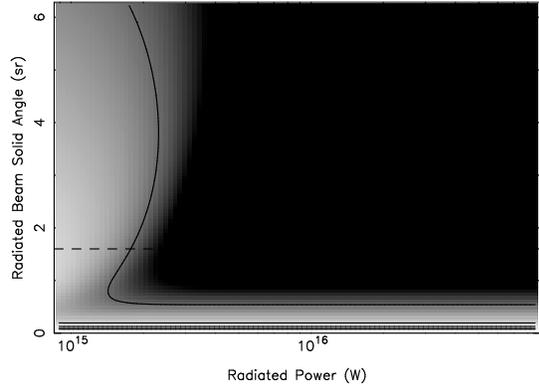}}
\vspace*{-2ex}
\caption[]{As for Figure~\ref{fig:likeli}, but for a next generation,
long-wavelength instrument (e.g., the LWA or LOFAR) capable of flux
density measurements in the range of~25~mJy with an rms noise level
near~7~\mjybm.  In this case, five measurements were simulated.}
\label{fig:future}
\end{figure}

We have also considered the requirements on an instrument in order to
improve the constraints significantly.  Figure~\ref{fig:future}
illustrates the constraints placed by~5 measurements with an
instrument capable of obtaining flux density measurements\footnote{%
As in \cite{lfdghjh04}, we assume relatively short observation
durations.  Given the bursty characteristic of solar system planetary
radio emission, it is not clear that the typical radio astronomical
practice of conducting long integrations is appropriate.}
of approximately 25~mJy with rms noise levels of approximately
7~\mjybm.  Obtaining significant (an order of magnitude or better)
improvements on the powers radiated by extrasolar planetary electron
cyclotron masers will require measurements at this level.  One
significant advantage that future long-wavelength instruments are
likely to have over current instruments, though, is that future
instruments are likely to have a multi-beaming capability.
Consequently, there is likely to be significantly more time for
observation and a much larger number of measurements (or even
detections!) will be obtained.

\section{Conclusions}\label{sec:conclude}

We report three (3) epochs of 4-meter wavelength (74~MHz) observations
of \tauBoo\ with the \hbox{VLA}.  Our objective was to detect the
electron cyclotron maser emission from the planet orbiting this star,
assuming that the planet does produce the equivalent of Jovian radio
emissions.  In none of our 3 epochs did we detect emission at the
location of \tauBoo.  For a single epoch (Table~\ref{tab:limits}), our
limits on its emission range from~150 to~300~mJy, equivalent to a
range of luminosities of approximately 1--$2.5 \times
10^{16}$~\hbox{W}.  We have also stacked (``co-added'') the images to
produce an average limit of~165~mJy, equivalent to a luminosity of
$1.4 \times 10^{16}$~\hbox{W}.

We have developed a likelihood method to consider multi-epoch measurements.
Our likelihood function depends upon three planetary parameters, the
planet's radiated power level~$P_{\mathrm{rad}}$, the beaming solid
angle of the electron cyclotron maser~$\Omega$, and the characteristic
wavelength of emission~$\lambda_c$.  Under the assumption that the
planet does radiate at our observed wavelength, the parameters
$P_{\mathrm{rad}}$ and~$\Omega$ are degenerate
(Figure~\ref{fig:likeli}).  We find that the typical radiated power
must be less than about~$10^{16}$~\hbox{W} unless the beaming solid
angle is $\Omega \ll 1$~sr, which would be considerably smaller than the
typical value for solar system planets.  Power levels
$P_{\mathrm{rad}} < 10^{16}$~W are within the predicted range
\citep{lfdghjh04,s04,gmmr05}, being comparable to or below the most recent
estimates that attempt to incorporate what is known about the star's
stellar wind strength.

We have also considered future long-wavelength instruments, such as
the LWA and \hbox{LOFAR}.  We find that to improve significantly upon
our constraints, future instruments will have to be able to measure
typical flux densities of approximately 25~mJy.

\acknowledgements
We thank P.~Zarka for illuminating discussions about solar system
planetary emissions, J.~Cordes for helpful discussions about
likelihood functions, students in the Thomas Jefferson High School for
Science and Technology astronomy lab for assistance with assessing the
utility of stacking, and the referee for several helpful comments
about the presentation and content.
The National Radio Astronomy Observatory is a facility of the National
Science Foundation operated under cooperative agreement by Associated
Universities, Inc.  Basic research in radio astronomy at the NRL is
supported by NRL 6.1 Base funding.


\begin{thebibliography}{}
\bibitem[\protect\citeauthoryear{Bastian, Dulk, \& Leblanc}{Bastian et
	al.}{2000}]{bdl00}
	Bastian, T.~S., Dulk, G.~A., \& Leblanc, Y.  2000, \apj, 545,
	1058

\bibitem[\protect\citeauthoryear{Butler et al.}{1997}]{bmwhs97}
	Butler, R.~P., Marcy, G.~W., Williams, E., Hauser, H.,
	\& Shirts, P.  1997, \apj, 474, L115
	%``Three New 51~Pegasi-Type Planets''

\bibitem[\protect\citeauthoryear{Catala et al.}{2007}]{cdsba07}
	Catala, C., Donati, J.-F., Shkolnik, E., Bohlender, D., \&
	Alecian, E.  2007, \mnras, 374, L42
	%The Magnetic Field of the Planet-Hosting Star $\tau$ Bootis

\bibitem[\protect\citeauthoryear{Cohen et al.}{2007}]{clk+07}
	Cohen, A.~S., Lane, W.~M., Kassim, N.~E., et al.  2007, \aj, in press

\bibitem[\protect\citeauthoryear{Condon et al.}{1998}]{ccgyptb98}
        Condon, J.~J., Cotton, W.~D., Greisen, E.~W., Yin, Q.~F.,
        Perley, R.~A., Taylor, G.~B., \& Broderick, J.~J.
        1998, \aj, 115, 1693

\bibitem[\protect\citeauthoryear{Cornwell \& Perley}{1992}]{cp92}
	Cornwell, T.~J.\ \& Perley, R.~A.  1992, \aap, 261, 353
	%Radio-interferometric Imaging of Very Large Fields - The Problem of Non-coplanar Arrays

\bibitem[\protect\citeauthoryear{Farrell, Desch, \& Zarka}{Farrell et al.}{1999}]{fdz99}  Farrell, W.~M., Desch, M.~D., \& 
	Zarka, P.  1999, \jgr, 104, 14025
%	``On the Possibility of Coherent Cyclotron Emission from
%	Extrasolar Planets'' 

\bibitem[\protect\citeauthoryear{Farrell et al.}{2003}]{fdlbz03}
	Farrell, W.~M., Desch, M.~D., Lazio, T.~J.~W., Bastian, T., \&
	Zarka, P.  2003, 
	%``Limits on the Magnetosphere/Stellar Wind
	%Interactions for the Extrasolar Planet about $\tau$~Boo,''
	in
	Scientific Frontiers in Research on Extrasolar Planets, eds.\
	D.~Deming \& S.~Seager (ASP: San Francisco) p.~151

\bibitem[\protect\citeauthoryear{Farrell et al.}{2004}]{fldbz04} 
	Farrell, W.~M., Lazio, T.~J.~W., Desch, M.~D., Bastian, T., \&
	Zarka, P.  2004,
	%``Radio Search for an Extrasolar Planet,''
	in
	Bioastronomy 2002: Life Among the Stars, eds.\ R.~Norris, C.~Oliver,
	\& F.~Stootman (ASP: San Francisco) p.~73

\bibitem[\protect\citeauthoryear{Green \& Gallagher}{1985}]{gg85}
	Green, J.~L., \& Gallagher, D.~L.  1985,
%	The Detailed Intensity Distribution of the AKR Emission Cone
	\jgr, 90, 9641

\bibitem[\protect\citeauthoryear{Griessmeier et al.}{2005}]{gmmr05}
        Griessmeier, J.-M., Motschmann, U., Mann, G., \& Rucker,
        H.~O.  2005, ``The Influence of Stellar Wind Conditions on the
        Detectability of Planetary Radio Emissions,'' Astron.\ \& Astrophys., 437, 717

\bibitem[\protect\citeauthoryear{Lazio et al.}{2004}]{lfdghjh04}
	Lazio, T.~J.~W., Farrell, W.~M., Dietrick, J., Greenlees, E.,
	Hogan, E., Jones, C., \& Hennig, L.~A.  2004, \apj, 612, 511

\bibitem[\protect\citeauthoryear{Majid et al.}{2006}]{mwcklnz06}
	Majid, W., Winterhalter, D., Chandra, I., Kuiper, T., Lazio,
	J., Naudet, C., \& Zarka, P.  2006, 
	%Search for Radio Emission from Extrasolar Planets:
	%Preliminary Analysis of GMRT Data
	in Planetary Radio Emissions~\hbox{VI}, eds.\ H.~O.~Rucker,
	W.~S.~Kurth, \& G.~Mann (Austrian Academy Science: Vienna) p.~589

\bibitem[\protect\citeauthoryear{Perryman et al.}{1997}]{plk+97}
	Perryman, M.~A.~C., Lindegren, L., Kovalevsky, J., et al.  1997, \aap, 323, L49

\bibitem[\protect\citeauthoryear{Ryabov et al.}{2004}]{rzr04}
	Ryabov, V.~B., Zarka, P., Ryabov, B.~P.  2004, Planet.\ Space
	Sci., 52, 1479--1491
	%Search of Exoplanetary Radio Signals in the Presence of
	%Strong Interference: Enhancing Sensitivity by Data
	%Accumulation

\bibitem[\protect\citeauthoryear{Shkolnik et al.}{2005}]{swbgk05}
	Shkolnik, E., Walker, G.~A.~H., Bohlender, D.~A., Gu, P.-G., 
	\& Kuerster, M.  2005, \apj, 622, 1075
	%Hot Jupiters and Hot Spots: The Short- and Long-term Chromospheric Activity on Stars with Giant Planets

\bibitem[\protect\citeauthoryear{Stevens}{2005}]{s04}
	Stevens, I.~R.  2005, \mnras, 356, 1053
%	``Magnetospheric Radio Emission from Extrasolar Giant Planets: The Role of the Host Stars''

\bibitem[\protect\citeauthoryear{Winglee, Dulk, \& Bastian}{Winglee et
	al.}{1986}]{wdb86}
	Winglee, R.~M., Dulk, G.~A., \& Bastian, T.~S.  1986, \apj,
	309, L59

\bibitem[\protect\citeauthoryear{Wood et al.}{2005}]{wmzlr05}
	Wood, B.~E., Mueller, H.-R., Zank, G.~P., Linsky, J.~L., \&
	Redfield, S.  2005, \apj, 628, L143

\bibitem[\protect\citeauthoryear{Worsley et al.}{2005}]{wfb+05}
	Worsley, M.~A., Fabian, A.~C., Bauer, F.~E., et al.  2005,
	%``The Unresolved Hard X-ray Background: The Missing Source
	%Population Implied by the Chandra and XMM-Newton Deep
	%Fields,''
	\mnras, 357, 1281

\bibitem[\protect\citeauthoryear{Yantis, Sullivan, \& Erickson}{Yantis et al.}{1977}]{yse77}
	Yantis, W.~F., Sullivan, W.~T., \hbox{III}, \& Erickson,
	W.~C.  1977, Bull.\ Amer.\ Astron.\ Soc., 9, 453

\bibitem[\protect\citeauthoryear{Zarka}{2007}]{z07}
	Zarka, P.  2007, 
	%``Plasma Interactions of Exoplanets with their Parent Star 
	%and Associated Radio Emissions''
	Planet.\ Space Sci., in press

\bibitem[\protect\citeauthoryear{Zarka}{2006}]{z06}
	Zarka, P.  2006, in Planetary Radio Emission~\hbox{VI}, eds.\ 
	H.~O.~Rucker et al.\ (Austrian Acad.: Vienna) p.~543

\bibitem[\protect\citeauthoryear{Zarka \& Cecconi}{2004}]{zc04}
	Zarka, P., \& Cecconi, B.  2004, \jgr, 109, A09S15

\bibitem[\protect\citeauthoryear{Zarka et al.}{2001}]{ztrr01}
	Zarka, P., Treumann, R.~A., Ryabov, B.~P., \& Ryabov, V.~B.
	2001, Ap\&SS, 277, 293 
%	``Magnetically-Driven Planetary Radio Emissions and
% 	Application to Extrasolar Planets,''

\bibitem[\protect\citeauthoryear{Zarka et al.}{1997}]{zqr+97}
	Zarka, P., Queinnec, J., Ryabov, B.~P., et al.  1997, in
	Planetary Radio Emission~\hbox{VI},  eds.\ H.~O.~Rucker et
	al.\ (Austrian Acad.: Vienna) p.~101

\bibitem[\protect\citeauthoryear{Zibetti et al.}{2005}]{zwsb05}
	Zibetti, S., White, S.~D.~M., Schneider, D.~P., \& Brinkmann,
	J.  2005, 
	%``Intergalactic Stars in $z \sim 0.25$ Galaxy
	%Clusters: Systematic Properties from Stacking of Sloan Digital
	%Sky Survey Imaging Data,''
	\mnras, 358, 949
\end{thebibliography}
\end{document}